\documentclass{appolb}
\usepackage{graphicx}
\newcommand \bea {\begin{eqnarray} }
\newcommand \eea {\end{eqnarray}}

\begin{document}
\title{Susy atomic model%
\thanks{Presented at the Strongly Correlated Electron Systems 
Conference, Krak\'ow 2002. This work was supported by NSF grant DMR 9983156. Thanks to C. P{\'e}pin and O. Parcollet for discussions. }%
}


\author{J.Hopkinson and P.Coleman
\address{Center for Materials Theory, Department of Physics and Astronomy, Rutgers University, 136 Frelinghuysen Rd., Piscataway, NJ, 08854, USA.}
}
\maketitle


\begin{abstract}
We present the simplest model to which one can apply the supersymmetric Hubbard operators recently introduced.\cite{cit1}  For the atomic model, $H=-E_d X_{00}$, where $X_{00} = |0><0|$ is a Hubbard operator and $E_d$ is the energy of the localized spin level, we show how one can develop exact solutions for the entropy and heat capacity as a function of temperature.  With this gold standard we are able to develop a controlled approximation scheme to field theoretically treat the susy approximation at the level of mean field + gaussian corrections and test its accuracy against the widely used slave boson and slave fermion approximations.  We find that in addition to slave boson and slave fermion solutions, a new class of solutions exists in the physical case Q = 1, N = 2 which can be properly treated by neither previously existing approach.  The phase diagram generated by the mean field saddle-point bears a superficial resemblance to the V-shaped phase diagram common to systems close to a quantum critical point and may provide a natural starting point for investigations of strongly correlated models capturing this physics.
\end{abstract}


  
\section{Introduction}
In heavy fermion systems close to an antiferromagnetic quantum critical point (aqcp) one now has the ability to tune from an antiferromagnetic state to a paramagnetic state as a function of field\cite{yb}, pressure\cite{yb,ce} or doping\cite{yb,ce}.  In the region above such a point, anomalous behavior is seen in measurements of heat capacity $\frac{C_v}{T} \approx \ln(\frac{T_0}{T})$ and resistivity $\rho \approx T$.  The question has arisen in the literature whether these exponents are due to a dynamical d=2 local qcp\cite{si} or rather a symptom of a breakdown of the d=3 Fermi liquid\cite{coleman}.  In particular, the observation of hyperscaling in these compounds has been used to argue the need for a microscopic theory capable of describing the breakdown of the Fermi surface and the formation of antiferromagnetic order as one tunes through this point.

Susy Hubbard operators\cite{cit1} with SU(1$|$1)xU(1) symmetry allow one to tune from a bosonic description of a spin (magnetic) to a fermionic description (Fermi surface).  For an atomic model, we show using counting arguments how it is possible to generate exact solutions for the entropy and heat capacity as a function of temperature for the Hilbert space of these operators.  Comparison of these with field theoretic results (mean field + gaussian fluctuations) allows us to estimate the error endemic to this $\frac{1}{N}$ approach, contrasting it with slave fermion and slave boson approaches.

\section{An exact solution}

To fix an irreducible representation of the Hubbard operators, we set 
\bea
|a><b| = X_{ab} = B^{\dagger}_a B_b + F^{\dagger}_a F_b
\eea
where $F_a = (f_1,...,f_N, \phi)$ and $B_a = (b_1,...,b_N, \chi)$ defines spin fields to be fermionic or bosonic respectively and their slave partners the converse, while maintaining the constraints
$Q = n_{b}
+ n_{ \phi} + n_{f} + n_{\chi}$ and
$Y = n_{\phi} + n_{f} - (n_{ b} + n_{\chi}) + \frac{1}{Q}[\theta, \theta^{\dagger}]$
where
$\theta = \sum_{\sigma}{b}_{\sigma }^{\dagger}{f}_{\sigma } - {\chi}^{\dagger}{\phi}$ is an operator interconverting fermions and bosons for the corner state.  This generates a series of L-shaped Young tableaux, the simplest of which (a single box) corresponds to a single physical spin when N=2.  To find the exact free energy of the state (Q,Y) we simply count the number of available states to the system at a given energy level.  Defining $h = \frac{Q+Y+1}{2}$ and $w=\frac{Q-Y+1}{2}$, the number of states available are:
\bea
\lambda_N(h,w) =\left(\begin{array}{c}N \\ h \end{array}\right)\left(\begin{array}{c}N + w - 1 \\ w \end{array}\right)\frac{wh}{N(w + h - 1)}
\eea
For an atomic model, the Hamiltonian is given by
\bea
H = E_dX_{\sigma \sigma} = E_d Q - E_dX_{00} = -E_dX_{00}
\eea
where $E_d$ is the energy of the d or f-electron state (we have dropped a constant in the free energy), which leads to a partition function of the form
\bea
Z = \sum_{i=0}^{h-1} \sum_{j=0}^{1} \lambda_N(h-i,w-j)e^{\beta (i+j)E_d}
\eea
and the free energy $F=-T\ln(Z)$, entropy $S=-\partial_TF$ and heat capacity $C_v = T\partial_TS$ immediately follow.
Including constraints, at the mean field level the Hamiltonian becomes
\bea
H = &-&E_d (\hat n_{\phi} + \hat n_{\chi}) + \lambda(\hat n_b + \hat n_f + \hat n_{\chi} + \hat n_{\phi} - Q_0) \nonumber \\ &+& \zeta (\hat n_f + \hat n_{\phi} - (\hat n_b + \hat n_{\chi}) + \frac{1}{Q_0}<[\theta,\theta^{\dagger}]> - Y_0)
\eea
where we can only evaluate the last term at the level of gaussian fluctuations.  Nonetheless we include it in the saddle-point, as we additionally treat the effects of gaussian fluctuations in the bosonic/fermionic character,
\bea
F = N F_f + N F_b + F_{\chi} + F_{\phi}+ F_{\eta} -\lambda Q_0 - \zeta Y_0 +  F_{\delta \lambda_f} + F_{\delta \lambda_b}
\eea
which at the mean field level sets
\bea
&\tilde n_f + \frac{1}{N} n_{\phi} + \frac{1}{N} (1 - n_{\alpha}) = \tilde h \\ &\tilde n_b + \frac{1}{N}n_{\chi} + \frac{1}{N} n_{\alpha} = \tilde w 
\eea
where $n_{\alpha} = \frac{1}{e^{2\beta \zeta} + 1}$, $\tilde n_f =\frac{1}{e^{\beta (\lambda_f)} + 1}$, $\tilde n_b=\frac{1}{e^{\beta (\lambda_b)} - 1} $,$n_{\phi} = \frac{1}{e^{-\beta (E_d-\lambda_f)} - 1}$ and $n_{\chi} = \frac{1}{e^{-\beta (E_d - \lambda_b)} + 1}$.  Analytic solution of these equations leads to the phase diagrams shown in Fig.1 for the special case N=2.  For general I shaped Young tableaux one recovers the mean field results of slave fermions (vertical) and slave fermions (horizontal) although the free energy contains divergent terms in this limit.

\begin{figure}[!ht]
\begin{center}
\includegraphics[width=0.9\textwidth]{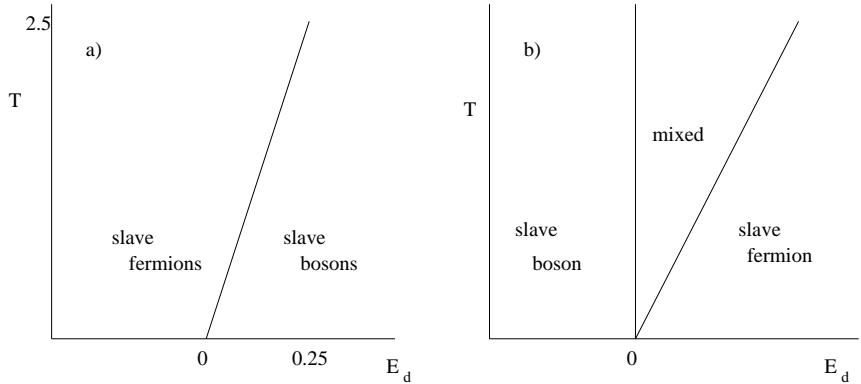}
\end{center}
\caption{The analytic saddle-point (s-p) solutions at N = 2 allow two possible phase diagrams (lines indicate equal free energies): a) The physical s-p picks the smallest free energy of the saddle-point solutions. b) A non-physical s-p which picks the highest free energy.  While minimization of the free energy does choose the correct physical result, this second solution illustrates how a non-trivial phase diagram may arise from a symmetry of the underlying formalism.  In an interacting model one might imagine that if the exact solution has $\frac{C_v}{T} \approx \ln(\frac{T_0}{T})$ then perhaps in that region a similar phase diagram to b) may be the energetically favourable one.}
\label{samplefig}
\end{figure}

While the appearance of a mixed phase in the phase diagram of Fig.1b is quite suggestive given that the mean field entropies also match along these lines, 
one is ultimately disappointed when one realizes that: i) although we expected that the magnetic phase would require a bosonic description of the spins--slave fermions and correspond to more tightly bound spins (Ed$<$0) and  slave bosons provide a natural candidate for a heavy Fermi surface which might be expected to appear close to (Ed=0) this does not seem to be the case here; ii) as the slave fermion mean field over-estimates the entropy at N=2, at the mean field level the mixed phase (interpolating to the slave boson mean field) ends up having a negative heat capacity.  This problem appears to be over-come by including the gaussian corrections, but in the limits $h\rightarrow1$ or $w\rightarrow1$ the gaussian fluctuations are not well-defined.  If after removal of the non-physical divergences the entropies of the mixed phase still match those of slave boson and slave fermion along the phase boundary lines (as was suggested by the mean field), then the positive difference $S_{slave boson}(T\rightarrow \infty) - S_{slave fermion}(T = \frac{d}{\ln[2]})$ would imply a small, positive heat capacity for this interval, in accordance with the exact result. iii) even were this the case, one would have to accept an ansatz choosing the maximal saddle-point free energy to admit Fig. 1b. 

In conclusion, we have shown that for a simple atomic model one recovers the mean field constraints known from slave boson and slave fermion treatments.  We have shown that a non-trivial mixed solution exists even in the physically relevant case Q=1, N=2.  Study of the properties of mixed solutions in the controllable large N limit may be of interest, and for the atomic model can be compared with exact results shown here.  A more thorough treatment will be given in the near future.{\cite{hopkinson}}



\begin{thebibliography}{}

\bibitem{cit1} P.Coleman, C.P{\'e}pin, J.Hopkinson, \textit{Phys. Rev. B} \textbf{62}, 3852 
(2000).

\bibitem{yb} Trovarelli et al. \textit{Phys. Rev. Lett.} {\textbf{85}}, 626, (2000).

\bibitem{ce} Schroeder et al. \textit{Nature} {\textbf{407}} 351, (2000).

\bibitem{si} Q.Si, S.Rabello, K.Ingersent, J.L.Smith, \textit{Nature},\textbf{413}, 804, (2001).

\bibitem{coleman} P. Coleman, Physica B, {\bf{259-261}}, 353, (1998).
\bibitem{hopkinson} J.Hopkinson, P.Coleman, cond-mat/0202060, to be published.
\end{thebibliography}
\end{document}